\documentclass[aps,prl,twocolumn,superscriptaddress,showpacs,preprintnumbers,amsmath,amssymb]{revtex4}

\usepackage{graphicx} % Include figure files
\usepackage{color}

\usepackage{dcolumn}  % Align table columns on decimal point

\graphicspath{{ps}}

\begin{document}

%\newcommand{\VERSION}{20040818a}

%%%%%%%%%%%%%%%%%%%%%%%%%%%%%%%%%%%%%%%%%%%%%%%%
% new commands %%%%%%%%%%%%%%%%%%%%%%%%%%%%%%%%%
%%%%%%%%%%%%%%%%%%%%%%%%%%%%%%%%%%%%%%%%%%%%%%%%

%======================
%units
%======================
\newcommand{\micron}{\mbox{~}\mu\mbox{m}}
\newcommand{\cm}{\mbox{~cm}}

\newcommand{\GeV}{\mbox{~GeV}}
\newcommand{\GeVc}{\GeV/c}
\newcommand{\GeVcc}{\GeVc^2}
\newcommand{\MeV}{\mbox{~MeV}}
\newcommand{\MeVc}{\MeV/c}
\newcommand{\MeVcc}{\MeVc^2}
\newcommand{\invfb}{\mbox{~fb}^{-1}}

\newcommand{\ee}{e^{+}e^{-}}
\newcommand{\ks}{K^0_{S}}

%==========================
% B -> K1 gamma
%==========================

\newcommand{\kpp}{K\pi^+\pi^-}
\newcommand{\kppp}{K^+\pi^+\pi^-}
\newcommand{\kzpp}{K^0\pi^+\pi^-}
\newcommand{\kspp}{K^0_{S}\pi^+\pi^-}
\newcommand{\kpppg}{\kppp\gamma}
\newcommand{\ksppg}{\kspp\gamma}
\newcommand{\mkpp}{M_{K^+\pi^+\pi^-} }
\newcommand{\krho}{K\rho}
\newcommand{\kstp}{K^*\pi}
\newcommand{\kzstp}{K_0^*\pi}
\newcommand{\konei}{K_1 (1270)}
\newcommand{\koneii}{K_1 (1400)}
\newcommand{\koneiz}{K_1 (1270)^0}
\newcommand{\koneiiz}{K_1 (1400)^0}
\newcommand{\koneip}{K_1 (1270)^+}
\newcommand{\koneiip}{K_1 (1400)^+}
\newcommand{\ktwost}{K_2^* (1430)}
\newcommand{\ktwostp}{K_2^* (1430)^+}
\newcommand{\ktwostz}{K_2^* (1430)^0}
\newcommand{\kres}{K_{\rm res}}
\newcommand{\kresp}{\kres^+}
\newcommand{\kppg}{K\pi^+\pi^-\gamma}
\newcommand{\kzppg}{K^0\pi^+\pi^-\gamma}
\newcommand{\btokppg}{B \rightarrow K\pi^+\pi^-\gamma}
\newcommand{\bptokpppg}{B^+ \rightarrow K^+\pi^+\pi^-\gamma}
\newcommand{\bztokzppg}{B^0 \rightarrow K^0\pi^+\pi^-\gamma}

\newcommand{\koneig}{\konei\gamma}
\newcommand{\koneiig}{\koneii\gamma}
\newcommand{\ktwostg}{\ktwost\gamma}

\newcommand{\koneipg}{\koneip\gamma}
\newcommand{\koneiipg}{\koneiip\gamma}
\newcommand{\ktwostpg}{\ktwostp\gamma}

\newcommand{\btokoneg}{B \to K_1 \gamma}
\newcommand{\btokoneig}{B \rightarrow K_1 (1270) \gamma}
\newcommand{\bptokoneipg}{B^+ \rightarrow K_1(1270)^+ \gamma}
\newcommand{\bztokoneizg}{B^0 \rightarrow K_1(1270)^0 \gamma}
\newcommand{\btokoneiig}{B \rightarrow K_1 (1400) \gamma}
\newcommand{\bptokoneiipg}{B^+ \rightarrow K_1 (1400)^+ \gamma}
\newcommand{\bztokoneiizg}{B^0 \rightarrow K_1 (1400)^0 \gamma}
\newcommand{\btoktwostg}{B \to K_2^* (1430)\gamma}
\newcommand{\bptoktwostpg}{B^+\to\ktwostpg}
\newcommand{\bptokpphig}{B^+ \to K^+\phi\gamma}
\newcommand{\btokresg}{B \to \kres\gamma}
\newcommand{\bptokrespg}{B^+ \to \kresp\gamma}
\newcommand{\mpp}{M_{\pi\pi}}
\newcommand{\mkp}{M_{K\pi}}

\newcommand{\btokstarg}{B \rightarrow K^*\gamma}
\newcommand{\btos}{b \rightarrow s}
\newcommand{\btosgam}{b \rightarrow s\gamma}
\newcommand{\BBbar}{B\overline{B}}
\newcommand{\qqbar}{q\overline{q}}
\newcommand{\Dbar}{\overline{D}{}}

%==========================
% variables
%==========================

\newcommand{\Ebeam}{E^*_{\rm beam}}
\newcommand{\Egamma}{E^*_\gamma}
\newcommand{\Ekpp}{E^*_{K\pi\pi}}
\newcommand{\thetagamma}{\theta_\gamma}
\newcommand{\pcmvec}{\vec{p}^{\;*}}
\newcommand{\pivec}{\vec{p}_i^{\;*}}
\newcommand{\pjvec}{\vec{p}_j^{\;*}}
\newcommand{\pmvec}{\vec{p}_m^{\;*}}
\newcommand{\thetaij}{\theta_{ij}}
\newcommand{\thetaim}{\theta_{im}}
\newcommand{\Qijl}{Q_{ij,l}}
\newcommand{\abs}[1]{\left|#1\right|}
\newcommand{\mbc}{M_{\rm bc}}
\newcommand{\de}{\Delta E}
\newcommand{\lrcpp}{LR(cos\theta_{\pi\pi})}

\newcommand{\pBvec}{\pcmvec_B}
\newcommand{\pgamvec}{\pcmvec_\gamma}
\newcommand{\pkppvec}{\pcmvec_{K\pi\pi}}
\newcommand{\Br}{{\cal B}}
\newcommand{\calL}{{\cal L}}
\newcommand{\calR}{{\cal R}}

%==========================
% journals
%==========================

\def\Journal#1#2#3#4{{#1} {\bf #2}, #3 (#4)}
\def\NCA{Nuovo Cimento}
\def\NIMA{Nucl. Instrum. Meth. A}
\def\NPB{Nucl. Phys. B}
\def\PLB{Phys. Lett. B}
\def\PRL{Phys. Rev. Lett.}
\def\PRD{Phys. Rev. D}
\def\ZPC{Z. Phys. C}
\def\EPJD{Eur. Phys. J. direct C}
\def\EPJC{Eur. Phys. J. C}
\def\etal{{\it et al.}}

%==========================
% results
%==========================

\newcommand{\EM}[1]{\times10^{-#1}}
\newcommand{\PM}[2]{^{+#1}_{-#2}{}}
\newcommand{\PlMi}[2]{(+#1/{-}#2)}

\newcommand{\Nkpppg}{318\pm22}   
\newcommand{\Nkzppg}{67\pm10}   
\newcommand{\Ekpppg}{ (8.36\pm 0.64)\%}
\newcommand{\Ekzppg}{ (1.82\pm 0.12)\%}
\newcommand{\Bkpppg}{ (2.50\pm{0.18}\pm{0.22})\EM5 }
\newcommand{\Bkzppg}{ (2.4\pm0.4\pm0.3)\EM5 } 
\newcommand{\Skpppg}{16\sigma}
\newcommand{\Skzppg}{8.3\sigma}

\newcommand{\NkoneigI}{   102 \pm 22 } 
\newcommand{\EkoneigI}{   (1.56\pm0.12)\% }
\newcommand{\EkppgnrI}{   (3.80\pm0.29)\% }
\newcommand{\BkoneigIfull}{  
         (4.3\pm0.9{\rm(stat.)}\pm0.9{\rm(syst.)})\EM5 }

\newcommand{\BkoneigI}{  (4.3\pm0.9\pm0.9)\EM5 }
\newcommand{\BkoneiigI}{ <2.95\EM5 }
\newcommand{\BkppgnrI}{  (1.79\PM{0.74}{0.70}\PM{0.44}{0.45})\EM5 }
\newcommand{\SkoneigI}{  7.3\sigma }

\newcommand{\NkoneigII}{    94.9\PM{22.4}{20.5} }
\newcommand{\NkoneiigII}{   23 \pm{26} } 
\newcommand{\EkoneiigII}{  (2.68\pm0.20)\% }
\newcommand{\BkoneiigII}{ <1.5\EM5 }

\newcommand{\NkoneigZI}{    15.0\PM{10.1}{8.2} }
\newcommand{\EkoneigZI}{   (0.34\pm0.02)\% }
\newcommand{\BkoneigZI}{  <5.8\EM5 }

\newcommand{\NkoneiigZII}{   -7.8\PM{16.3}{13.8} }
\newcommand{\EkoneiigZII}{  (0.61\pm0.04)\% }
\newcommand{\BkoneiigZII}{ <1.2\EM5 }

%%%%%%%%%%%%%%%%%%%%%%%%%%%%%%%%%%%%%%%%%%%%%%%%
% end new commands %%%%%%%%%%%%%%%%%%%%%%%%%%%%%
%%%%%%%%%%%%%%%%%%%%%%%%%%%%%%%%%%%%%%%%%%%%%%%%

\title{ \quad\\[0.5cm] \boldmath{ Observation of $\bptokoneipg$ } }

%%%%%%%%%%%%%%%%%%%%%%%%%%%%%%%%%%%%%%%%%%%%%%%%%%%%%%%%%%%%%%%%%%%%%
%%%% >>>>> insert the authorlist here. BEFORE the abstract !!!!! <<<<<
%%%%%%%%%%%%%%%%%%%%%%%%%%%%%%%%%%%%%%%%%%%%%%%%%%%%%%%%%%%%%%%%%%%%%

\affiliation{Budker Institute of Nuclear Physics, Novosibirsk}
\affiliation{Chiba University, Chiba}
\affiliation{Chonnam National University, Kwangju}
\affiliation{University of Cincinnati, Cincinnati, Ohio 45221}
\affiliation{University of Frankfurt, Frankfurt}
\affiliation{Gyeongsang National University, Chinju}
\affiliation{University of Hawaii, Honolulu, Hawaii 96822}
\affiliation{High Energy Accelerator Research Organization (KEK), Tsukuba}
\affiliation{Hiroshima Institute of Technology, Hiroshima}
\affiliation{Institute of High Energy Physics, Chinese Academy of Sciences, Beijing}
\affiliation{Institute of High Energy Physics, Vienna}
\affiliation{Institute for Theoretical and Experimental Physics, Moscow}
\affiliation{J. Stefan Institute, Ljubljana}
\affiliation{Kanagawa University, Yokohama}
\affiliation{Korea University, Seoul}
\affiliation{Kyungpook National University, Taegu}
\affiliation{Swiss Federal Institute of Technology of Lausanne, EPFL, Lausanne}
\affiliation{University of Ljubljana, Ljubljana}
\affiliation{University of Maribor, Maribor}
\affiliation{University of Melbourne, Victoria}
\affiliation{Nagoya University, Nagoya}
\affiliation{Nara Women's University, Nara}
\affiliation{National Central University, Chung-li}
\affiliation{National United University, Miao Li}
\affiliation{Department of Physics, National Taiwan University, Taipei}
\affiliation{H. Niewodniczanski Institute of Nuclear Physics, Krakow}
\affiliation{Nihon Dental College, Niigata}
\affiliation{Niigata University, Niigata}
\affiliation{Osaka City University, Osaka}
\affiliation{Osaka University, Osaka}
\affiliation{Panjab University, Chandigarh}
\affiliation{Peking University, Beijing}
\affiliation{Princeton University, Princeton, New Jersey 08545}
\affiliation{Saga University, Saga}
\affiliation{University of Science and Technology of China, Hefei}
\affiliation{Seoul National University, Seoul}
\affiliation{Sungkyunkwan University, Suwon}
\affiliation{University of Sydney, Sydney NSW}
\affiliation{Tata Institute of Fundamental Research, Bombay}
\affiliation{Toho University, Funabashi}
\affiliation{Tohoku Gakuin University, Tagajo}
\affiliation{Tohoku University, Sendai}
\affiliation{Department of Physics, University of Tokyo, Tokyo}
\affiliation{Tokyo Institute of Technology, Tokyo}
\affiliation{Tokyo Metropolitan University, Tokyo}
\affiliation{Tokyo University of Agriculture and Technology, Tokyo}
\affiliation{University of Tsukuba, Tsukuba}
\affiliation{Virginia Polytechnic Institute and State University, Blacksburg, Virginia 24061}
\affiliation{Yonsei University, Seoul}
\author{Heyoung~Yang}\affiliation{Seoul National University, Seoul} % Seoul
\author{M.~Nakao}\affiliation{High Energy Accelerator Research Organization (KEK), Tsukuba} % KEK
\author{K.~Abe}\affiliation{High Energy Accelerator Research Organization (KEK), Tsukuba} % KEK
\author{K.~Abe}\affiliation{Tohoku Gakuin University, Tagajo} % TohokuGakuin
\author{H.~Aihara}\affiliation{Department of Physics, University of Tokyo, Tokyo} % Tokyo
\author{Y.~Asano}\affiliation{University of Tsukuba, Tsukuba} % Tsukuba
\author{T.~Aushev}\affiliation{Institute for Theoretical and Experimental Physics, Moscow} % ITEP
\author{S.~Bahinipati}\affiliation{University of Cincinnati, Cincinnati, Ohio 45221} % Cincinnati
\author{A.~M.~Bakich}\affiliation{University of Sydney, Sydney NSW} % Sydney
\author{Y.~Ban}\affiliation{Peking University, Beijing} % Peking
% ...
% Lausanne
\collaboration{The Belle Collaboration}

			   %%%%%%%%%%%%%%%%%%
			   %%%  Abstract  %%%
			   %%%%%%%%%%%%%%%%%%

\begin{abstract}
We report the observation of the radiative decay $\bptokoneipg$ using a
data sample of $140\invfb$ taken at the $\Upsilon(4S)$ resonance with
the Belle detector at the KEKB $\ee$ collider.  We find the branching
fraction to be $\Br(\bptokoneipg)=\BkoneigIfull$ with a 
significance of $\SkoneigI$.  We find no significant signal for
$\bptokoneiipg$ and set an upper limit $\Br(\bptokoneiipg)\BkoneiigII$ at
the 90\% confidence level.  We also measure inclusive branching
fractions for $\bptokpppg$ and $\bztokzppg$ in the mass range
$1\GeVcc<M_{K^{+(0)}\pi^+\pi^-}<2\GeVcc$.
\end{abstract}

\pacs{13.25.Hw, 14.40.Nd}

\maketitle

\tighten

{\renewcommand{\thefootnote}{\fnsymbol{footnote}}}
\setcounter{footnote}{0}

			 %%%%%%%%%%%%%%%%%%%%%%
			 %%%  Introduction  %%%
			 %%%%%%%%%%%%%%%%%%%%%%

Radiative $B$ decays that occur through the flavor changing neutral
current process $\btosgam$ have been one of the most sensitive probes 
in the
search for physics beyond the Standard Model (SM).  The first observed
exclusive radiative decay mode was $\btokstarg$~\cite{kstg,k892}.
The second mode was $\btoktwostg$, for which evidence was reported by
CLEO and Belle~\cite{kstg}.
Theoretical predictions for the branching fractions of the 
unobserved exclusive
decays can be found in Refs.~\cite{theor,theor2}. 
The modes $\btokoneig$ and $\btokoneiig$ 
($K_{1}\rightarrow K\pi\pi$) can be used to measure the photon helicity,
which may differ from the SM prediction in some models beyond the 
SM~\cite{gronau}. 
The neutral mode $\bztokoneizg$, $\konei^0\rightarrow\ks\rho^0$ would also be
useful to measure time-dependent $CP$ violation 
that may arise from new physics~\cite{ags}.

In this paper, we report the observation of $\bptokoneipg$, which
is the first radiative $B$ meson decay mode that involves an
axial-vector resonance.  We study radiative decays in the $\kpppg$ and
$\ksppg$ final states, where we search for resonant structure in the
$\kpp$ system~\cite{knote}.  We also report inclusive measurements of $\bptokpppg$
and $\bztokzppg$, and the results of a search for $\bptokoneiipg$.  The
analysis is based on a data sample of $140\invfb$ (152 million
$\BBbar$ pairs) taken at the
$\Upsilon(4S)$ resonance with the Belle detector at the KEKB
$e^+e^-$collider~\cite{KEKB}.  

The Belle detector 
consists of a three-layer silicon vertex detector (SVD), a 50-layer
central drift chamber (CDC), an array of aerogel threshold \v{C}erenkov
counters (ACC), a barrel-like arrangement of time-of-flight
scintillation counters (TOF), and an electromagnetic calorimeter
comprised of CsI(Tl) crystals (ECL) located inside a superconducting
solenoid coil that provides a 1.5~T magnetic field.  
An instrumented iron flux-return
(KLM) for $K_L^0$ and muon identification is located outside of the coil.
The detector is described in detail
elsewhere~\cite{Belle}.

			%%%%%%%%%%%%%%%%%%%%%%%%
			%%%  Reconstruction  %%%
			%%%%%%%%%%%%%%%%%%%%%%%%

The photon candidate is the highest energy photon cluster measured with
the barrel ECL ($33^\circ < \theta_{\gamma} < 128^\circ$ in the
laboratory frame).
In order to reduce the background from
$\pi^0/\eta\to\gamma\gamma$ decays, we combine the photon candidate with all
other photon clusters in the event with energy greater than $30\MeV$
($200\MeV$) and reject the event if the invariant mass of any pair is
within $\pm18\MeVcc$ ($\pm32\MeVcc$) of the nominal $\pi^0$ ($\eta$)
mass.  
These correspond to $\pm 3 \sigma$ windows, where $\sigma$ is
the mass resolution.
We refer to this requirement as the $\pi^0/\eta$ veto.

Charged tracks
are required to have momentum in the center-of-mass (c.m.) frame greater
than $200\MeVc$ and to have an impact parameter relative to the
interaction point of less than $5\cm$ along the positron beam axis and
less than $0.5\cm$ 
in the transverse plane.
The charged kaon candidate is identified using a likelihood ratio
combining the information from the ACC, TOF, and CDC sub-detectors;
the remaining charged particles in the event are used as pion candidates,
unless the track has been identified as an electron, muon, or proton.

For neutral kaons, we use $\ks\to\pi^+\pi^-$ candidates that have
invariant masses within $\pm30\MeVcc$ of the nominal $\ks$ mass and a
c.m.\ momentum greater than $200\MeVc$.
The two pions are required to have a common vertex that is displaced
from the interaction point. The $\ks$ momentum direction is also
required to be consistent with the $\ks$ flight direction.

In order to study $B\to K_1\gamma$ ($K_1\to K\pi\pi$),
we first reconstruct $\btokppg$ inclusively,
without any requirement for the structure of the $K\pi^+\pi^-$
system.
We select $K\pi^+\pi^-$ combinations in the mass range
$1\GeVcc<M_{K\pi^+\pi^-}<2\GeVcc$.  
Given the $\kpp$ system and the photon candidate, 
we identify $B$ meson candidates using two
independent kinematic variables: the beam-energy constrained mass
$\mbc\equiv \sqrt{(\Ebeam/c^2)^2 - |\pBvec/c|^2 }$ and the energy
difference $\de\equiv \Ekpp + \Egamma - \Ebeam$, where $\Ebeam$ is the
beam energy and $\pBvec$ is the momentum of the $B$ candidate in the
c.m.\ frame~\cite{asterisk}.
The $B$ momentum is calculated as
$\pBvec = \pkppvec + \frac{\pgamvec}{\Egamma} \times (\Ebeam - \Ekpp)$
in order to improve the $\mbc$ resolution.
We select $B$ candidates within $-0.1\GeV<\de<0.08\GeV$ and
$\mbc>5.2\GeVcc$.  If there exist multiple candidates, we choose the
candidate with the highest confidence level for the $\kppp$
vertex fit ($\pi^+\pi^-$ vertex in the neutral case).

			  %%%%%%%%%%%%%%%%%%%%
			  %%%  Background  %%%
			  %%%%%%%%%%%%%%%%%%%%

The dominant background comes from hadronic continuum
($\ee$$\to$ $q\bar{q}$, $q=u,d,s,c$).
To suppress this background,
we use two variables: the $B$ flight
direction ($\cos\theta_B^*$) and a Fisher discriminant~\cite{fisher}
built from a set of shape variables~\cite{KSFW}.  
For signal, the $B$ flight direction follows a
$1-\cos^2\theta_B^*$ distribution while that of $q\bar{q}$ is nearly
uniform.  The likelihood function 
$\calL_{S(B)}^{\cos\theta_B^*}$ is modeled as a $2^{\rm nd}$ 
($1^{\rm st}$) order
polynomial for the signal (continuum background) from MC samples.
For the shape variables, we use 16 modified Fox-Wolfram
moments~\cite{fox-wolfram} calculated for the following groups of particles:
1) particles that form the
signal candidate, 2) the remaining charged particles, 3) the remaining
neutral particles, and 4) a hypothetical particle for the missing
momentum of the event.  The Fisher discriminant is obtained from these
moments and the scalar sum of the transverse momentum.  The likelihood
function $\calL_{S(B)}^{\rm Fisher}$ is
modeled as a bifurcated Gaussian function both for the signal and 
the continuum background from MC samples.

These likelihood functions are then combined to form 
$\calR_S=\calL_S^{\cos\theta_B^*}\calL_S^{\rm Fisher}$/
($\calL_S^{\cos\theta_B^*}\calL_S^{\rm Fisher} +
\calL_B^{\cos\theta_B^*}\calL_B^{\rm Fisher}$).
We determine the $\calR_S$ requirement by maximizing
$N_S\left/\sqrt{N_S+N_B}\right.$, where $N_S$ and $N_B$ are the expected
number of the signal and background events, respectively,
in $\mbc>5.27 \GeVcc$.  For this
purpose, we use $\btokoneig$ and $\btokoneiig$ signal Monte Carlo (MC) 
simulated data, 
assuming all the $\btokoneg$ branching
fractions are $1\EM5$.  We find $\calR_S>0.9$ is the 
optimal requirement.
This requirement retains $47\%$ of the signal events
while rejecting $98\%$ of the continuum background events.

			%%%%%%%%%%%%%%%%%%%%%%%%
			%%%  Binned 1-D fit  %%%
			%%%%%%%%%%%%%%%%%%%%%%%%

The signal yields for 
$B\to K\pi^+\pi^-\gamma$ are extracted from a
binned maximum likelihood fit to the $\mbc$ distribution.
In addition to the continuum background, we
consider four $B$ decay background sources: known $B$
decays through the $b\to c$ transition (referred to as the $b\to c$
background), hadronic $B$ decays through the $b\to u$, $d$ or $s$
transitions (charmless background), $\btokstarg$ background,
and radiative $b\to s$ decays to final states other than $K^*\gamma$ and
$\kppg$ (other $\btosgam$ background).  To suppress the
$\btokstarg$ background, we reject the event if $\de$ and $\mbc$
calculated from either $K\pi\gamma$ combination satisfy 
$-0.2\GeV<\de<0.1\GeV$ and $\mbc>5.27\GeVcc$.
The signal $\mbc$ distributions are each modeled as a Gaussian
function; 
its width is fixed using a data sample of 
$B\to D(\to K\pi\pi)\pi$ decays,
treating the primary pion as a high energy photon.
The shapes of the background $\mbc$ distributions are determined
using large MC samples.
We find that the sum of the continuum and
$b\to c$ backgrounds is described by an ARGUS function~\cite{ARGUS},
a smooth functional form that has a kinematic threshold at half
of the center of mass energy.
Charmless decays, $B\to
K^*\gamma$ and other $\btosgam$ decays are modeled
as a sum of an ARGUS function and a Gaussian function.
The normalization of the continuum plus $b\to c$ background
is floated in the fit; the normalization of the other three
components are fixed in the fit.

\begin{figure}[b]
\includegraphics[width=0.5\textwidth]{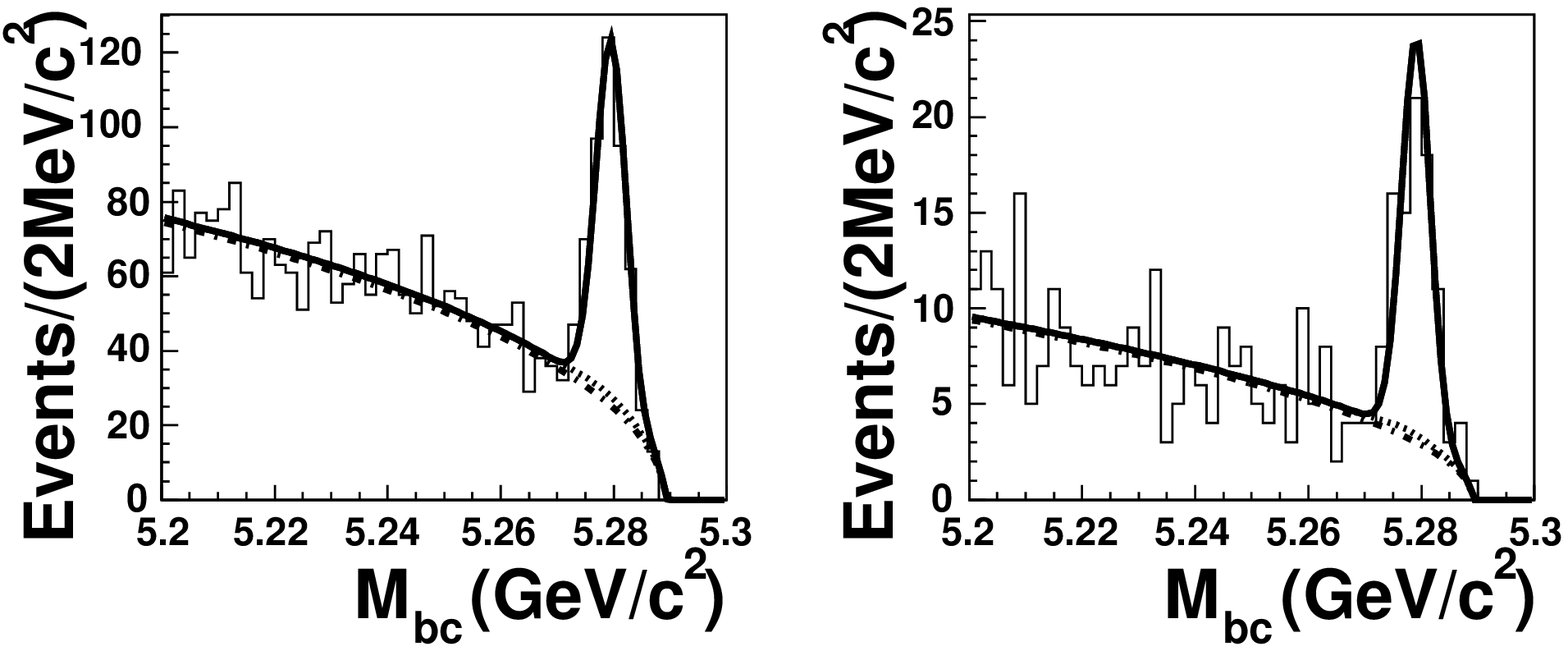}
\caption{$\mbc$ distributions for $\bptokpppg$ (left) and $\bztokzppg$
         (right). Curves show the continuum plus $b\to c$ background
         component (dot-dashed), total background (dotted) and the total
         fit result (solid). }
\label{mbc}
\end{figure}

The fit result is shown in Fig.~\ref{mbc}.  
For the $\bptokpppg$ mode, we
obtain $\Nkpppg$ events with a significance of $\Skpppg$,
where the significance is defined as $\sqrt{-2\ln({\calL}_0/{\calL}_{\rm
max})}$, and $\calL_{\rm max}$ and $\calL_0$ denote the maximum
likelihoods of the fit with and without the signal component,
respectively, and the significance includes systematic error.  
Similarly, we obtain $\Nkzppg$ events with a 
significance of $\Skzppg$ for the $\bztokzppg$ mode.

%%%%%%%%%%%%%%%%%%
%%%%%%%%%%%%%%%%%%

		 %%%%%%%%%%%%%%%%%%%%%%%%%%%%%%%%%%%%%%
		 %%% 1-D Fitting systematic errors %%%
		 %%%%%%%%%%%%%%%%%%%%%%%%%%%%%%%%%%%%%%

The systematic uncertainty related to the fitting procedure is
estimated in the following way.
We vary the width and the mean of the signal Gaussian by
the error of the $B\to D\pi$ calibration sample.
We vary the ARGUS parameter of the continuum plus $b\to c$ background
by the errors from fits to the MC sample and
to a data sideband region defined as $0.1\GeV<\de<0.5\GeV$,
then we take the quadratic sum of those errors.
The $\btokstarg$ component is varied by
the branching fraction uncertainty.  
The normalization of the other $\btosgam$ background component is
varied within its respective uncertainty, estimated from the uncertainties
in the total $\btosgam$ branching fraction \cite{HFAG} and the
fraction of $\kppg$ in the $s\gamma$ final state \cite{acpxsgam}.
For the charmless
background we vary the normalization by $\pm100\%$.  
We also assign the uncertainty due to a possible fit bias as 
the error of the fit to the signal MC sample.
The total fitting errors are $5.3\%$ ($12\%$) for the $\bptokpppg$
($\bztokzppg$) mode.

		       %%%%%%%%%%%%%%%%%%%%%%%%%%
		       %%%  Unbinned 2-D fit  %%%
		       %%%%%%%%%%%%%%%%%%%%%%%%%%

In order to decompose intermediate resonances that may be involved in
the $\kppp$ final state, we perform an unbinned maximum likelihood fit to
the $\mbc$ and $\mkpp$ distributions of the $\bptokpppg$ candidates.
There are
many possible resonances that can contribute: $\konei$, $\koneii$,
$\ktwost$, $K^*(1410)$, $K^*(1680)$, and so on.  We consider the
first three resonances, and include an additional non-resonant
$\bptokpppg$ component. 
The $\btoktwostg$ component, which is
already measured, is fixed in the fit.  
We model the $\mkpp$ distribution of the $\konei$ resonance
as a sum of three decay chains, $\konei^+\to K^+\rho^0$,
$\rho^0\to\pi^+\pi^-$; $\konei^+\to K^{*0}\pi^+$, $K^{*0}\to K^+\pi^-$;
and $\konei^+\to K_0^*(1430)^0\pi^+$, $K_0^*(1430)^0\to K^+\pi^-$. 
The $\mkpp$
distribution for each decay chain is described by convolving the two
relativistic Breit-Wigner functions of the resonances in the 
decay chain.  The $\mkpp$ distribution of the $\koneii$
resonance is modeled with a single decay chain, $\koneii^+\to
K^{*0}\pi^+$, $K^{*0}\to K^+\pi^-$.  The $\mkpp$ distributions of other
$\btosgam$, 
non-resonant $\bptokpppg$
and continuum plus $b\to c$ backgrounds are modeled using
the function $(p_0 + p_1 x) \exp(p_2 + p_3 x + p_4 x^2)$, where $x=\mkpp$,
and $p_i$ $(i=0\ldots4)$ are empirical parameters that are determined 
from MC samples.

%%%%%%%%%%%%%%%%%%
\begin{figure}[b]
\includegraphics[width=0.5\textwidth]{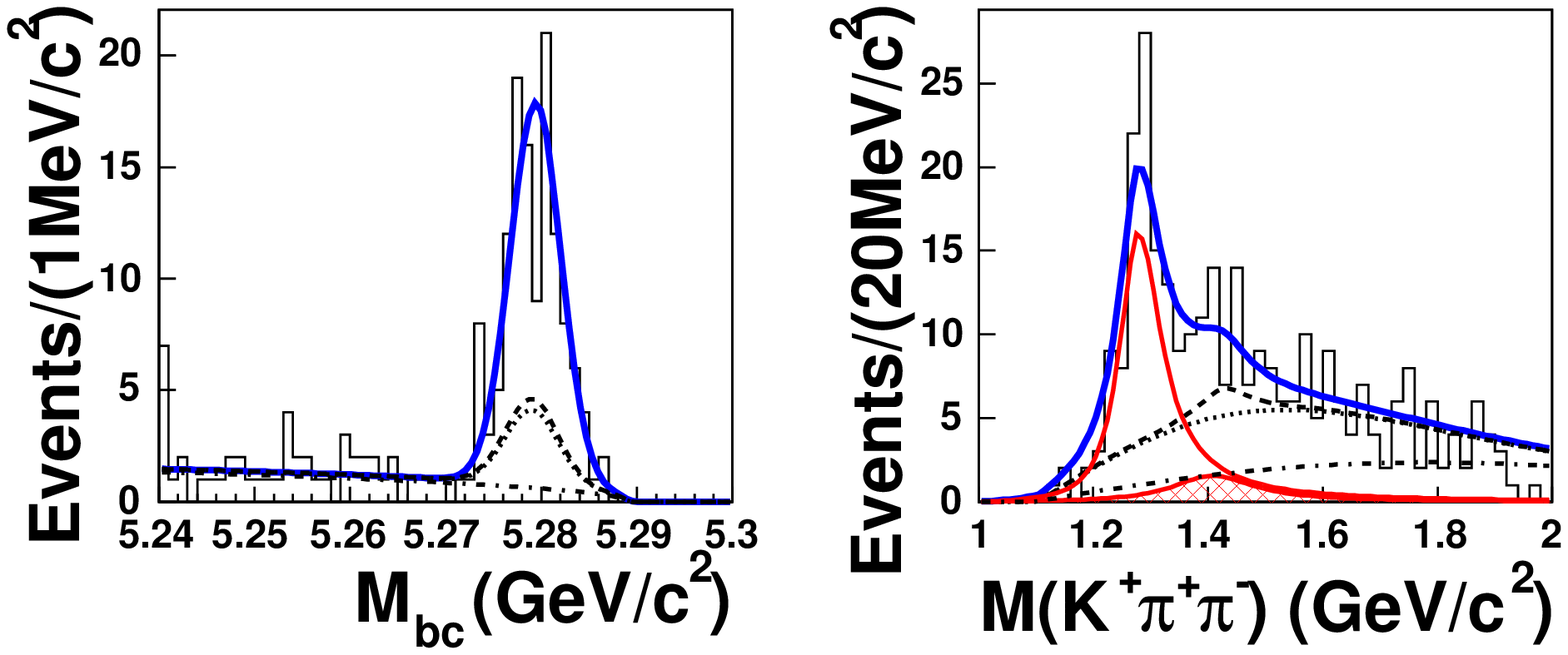}
\caption{$\mbc$ distribution for $1.2\GeVcc<\mkpp<1.4\GeVcc$ (left) and
          $\mkpp$ distribution for $\mbc>5.27\GeVcc$ (right) of the
          $\koneipg$ enriched sample with $0.6\GeVcc<\mpp<0.9\GeVcc$.
          Curves show the projections of the fit results for the
          continuum plus $b\to c$ background component (dot-dashed),
          total background without and with the $\ktwostpg$ component
          (dotted and dashed, respectively), $\koneipg$ (thin solid line)
          and $\koneiipg$ (hatched) components, and the sum of
          all components (thick solid line). }
\label{fig:k1g}
\end{figure}
%%%%%%%%%%%%%%%%%%

In order to enhance the $\konei$ component, we select events
with $\pi^+\pi^-$ mass in the $\rho^0$ mass region,
$0.6\GeVcc<\mpp<0.9\GeVcc$ ($\Br(\konei\to K\rho)$ = 
$(42\pm6)\%$ being much larger than 
$\Br(\koneii\to K\rho)$ = 
$(3\pm3)\%$~\cite{PDG}).  
The fit result
is shown in Fig.~\ref{fig:k1g}.  We find $\NkoneigI$ events for the
$\bptokoneipg$ component
with a significance of $\SkoneigI$,
while we fix the yields of the $K^*\gamma$, other $b\to s\gamma$
and charmless background components to be 2.3, 10.2 and 2.0 
events, respectively.
Similarly, we select the events with the $K^+\pi^-$ mass in the $K^{*0}$
mass region, $0.8\GeVcc<\mkp<1.0\GeVcc$, in order to enhance the
$\koneii$ component ($\Br(\koneii\to K^*\pi)$ =
$(94\pm6)\%$ being much larger than $\Br(\konei\to K^*\pi)$ =
$(16\pm5)\%$~\cite{PDG}).  
We find $\NkoneiigII$ events for
the $\bptokoneiipg$ component. 
Since the $\koneiipg$
component is not significant, we set a 90\% confidence level upper limit
on the signal yield, $N_{90}$, which is calculated from the relation
$\int_0^{N_{90}} \calL(n)dn=0.9\int_0^\infty \calL(n)dn$, where
$\calL(n)$ is the likelihood function with the signal yield fixed at
$n$.

		 %%%%%%%%%%%%%%%%%%%%%%%%%%%%%%%%%%%%%%
		 %%% 2-D Fitting systematic errors %%%
		 %%%%%%%%%%%%%%%%%%%%%%%%%%%%%%%%%%%%%%

The systematic uncertainty related to the fitting procedure includes
the same contributions as for the $K^+\pi^+\pi^-$ fit. 
The sub-branching fractions for the $\konei$ and the
branching fraction for the $\ktwostg$ as well the uncertainties
in these quantities are taken from Ref.~\cite{PDG} and
included in the fitting procedure.
We also assign the uncertainty due to incorrectly reconstructed
signal events, which is determined from a MC study.
The total unbinned fitting errors are 
$20\%$ ($32\%$) for the $\bptokoneipg$
($\bptokoneiipg$) mode.

			  %%%%%%%%%%%%%%%%%%%%
			  %%% Efficiencies %%%
			  %%%%%%%%%%%%%%%%%%%%

The reconstruction efficiency is obtained from the signal MC samples.
For the inclusive measurement, 
we calculate the efficiency using the $\konei^+\gamma$ and 
$K^+\pi^+\pi^-\gamma$ non-resonant MC samples, and weighting
according to the measured ratio of yields in those modes
($\koneiipg$ and $\ktwostpg$
contributions are neglected).  
We assume the same ratio for the $\bztokzppg$
efficiency.  
We estimate the systematic errors due to photon detection (2.8\%),
tracking (1\% per track), charged particle identification (0.5\% per particle),
$\ks$ reconstruction (4.5\%) and the likelihood ratio and $\pi^0/\eta$ veto 
(6.1\% for $\kpppg$, 3.2\% for $\kzppg$).
The total efficiency error is 7.6\% (6.6\%)
for the $\bptokpppg$ ($\bztokzppg$) mode.

		      %%%%%%%%%%%%%%%%%%%%%%%%%%%
		      %%% Branching fractions %%%
		      %%%%%%%%%%%%%%%%%%%%%%%%%%%

Using the signal yield and efficiency we find
\begin{equation}
\Br(\bptokoneipg) = \BkoneigI
\end{equation}
where the first (second) error is statistical (systematic)
assuming that the production of $B^+$ and $B^0$ 
in $\Upsilon(4S)$ decays is equal. 
We also measure the inclusive branching fractions of 
$B\to K\pi^+\pi^-\gamma$
given in Table~\ref{table:results}. We find that
$\Br(\bptokpppg)$ is consistent with the previous measurement 
with a significantly improved error~\cite{kxgam};
the neutral one is measured with a similar branching fraction.

Similarly, we perform an unbinned maximum likelihood fit 
to the $\bztokzppg$ candidates to decompose 
$\koneiz$ and $\koneiiz$. Due to the limited statistics we 
do not obtain a significant result 
and set only upper limits.

			   %%%%%%%%%%%%%%%%%%
			   %%% Conclusion %%%
			   %%%%%%%%%%%%%%%%%%

To conclude, we observe a new radiative decay mode, $\bptokoneipg$, with a
branching fraction of $\BkoneigIfull$, which is larger than
theory predictions $(0.5\sim2.0)\times 10^{-5}$ ~\cite{theor,theor2}.
The rates for $\btokoneig$ and $\btokoneiig$ are sensitive to
the magnitude and sign of the 
$\konei-\koneii$ mixing angle. The large rate for
$\bptokoneipg$ compared to $\bptokoneiipg$ may be explained
by a positive mixing angle~\cite{theor2}.
This measurement of $\btokoneig$ shows that in the future
it will be possible to determine the photon helicity
using $B\to K_1 \gamma$, $K_1\to K\pi\pi$
and that time-dependent $CP$ violation using 
$B^0\to \konei^0\gamma$, $\konei^0\to K_S^0\rho^0$ decays
can also be studied.
We also measure similar branching fractions
for the inclusive decays $\bptokpppg$ and $\bztokzppg$.
The latter is measured for the first time.

			 %%%%%%%%%%%%%%%%%%%%%
			 %  Acknowledgments  %
			 %%%%%%%%%%%%%%%%%%%%%

We thank the KEKB group for the excellent operation of the
accelerator, the KEK Cryogenics group for the efficient
operation of the solenoid, and the KEK computer group and
the NII for valuable computing and Super-SINET network
support.  We acknowledge support from MEXT and JSPS (Japan);
ARC and DEST (Australia); NSFC (contract No.~10175071,
China); DST (India); the BK21 program of MOEHRD and the CHEP
SRC program of KOSEF (Korea); KBN (contract No.~2P03B 01324,
Poland); MIST (Russia); MESS (Slovenia); NSC and MOE
(Taiwan); and DOE (USA).

		      %%%%%%%%%%%%%%%%%%%%%%%%%%%
		      
%

\onecolumngrid

%%%%%%%%%%%%%%%%%
\begin{table}[t]
%\caption{Fit results}
\caption{Yields from fits, detection efficiencies, branching fractions
with statistical and systematic errors, and significances.
The $\konei$ and $\koneii$ are reconstructed as $K\pi^+\pi^-$
states which come through their sub-decay channels.}
\label{table:results}
\begin{tabular}{l@{~~~~}c@{~~~~}c@{~~~~}c@{~~~~}c}
\hline %----------------------------------------------------------------
\hline %----------------------------------------------------------------
 & yield & efficiency & branching fraction($\Br$) & significance \\
\hline %----------------------------------------------------------------
$\bptokoneipg$ & $\NkoneigI$ & $\EkoneigI$ & $\BkoneigI$ & $\SkoneigI$ \\
$\bztokoneizg$  & $\NkoneigZI$ & $\EkoneigZI$ & $\BkoneigZI$ & --- \\
\hline %----------------------------------------------------------------
$\bptokoneiipg$ &
            $\NkoneiigII$ & $\EkoneiigII$ & $\BkoneiigII$ & --- \\
$\bztokoneiizg$ &
            $\NkoneiigZII$ & $\EkoneiigZII$ & $\BkoneiigZII$ & --- \\
\hline %----------------------------------------------------------------
$\bptokpppg$ & ~$\Nkpppg$~ & ~$\Ekpppg$~ & ~$\Bkpppg$~ & $\Skpppg$ \\
$\bztokzppg$ & ~$\Nkzppg$~ & ~$\Ekzppg$~ & ~$\Bkzppg$~ & $\Skzppg$ \\
\hline %----------------------------------------------------------------
\hline %----------------------------------------------------------------
\end{tabular}
\end{table}
%%%%%%%%%%%%%%%%%

\twocolumngrid


\begin{thebibliography}{99}
\bibitem{k892}
%
We use $K^*$ to denote $K^*(892)$ throughout the paper.
\bibitem{knote}
%
The inclusion of charge conjugate modes is implied unless otherwise
stated, and
$K$ denotes both $K^+$ and $K_S^0$ 
throughout the paper.
\bibitem{asterisk}
%
Variables with an asterisk are calculated in the
c.m.\  frame.
\end{thebibliography}
\end{document}